\documentclass[twocolumn,showpacs,pre,nofootinbib]{revtex4}
\newcommand{\figurewidth}{\columnwidth}
\newcommand{\beq}{\begin{eqnarray}}
\newcommand{\eeq}{\end{eqnarray}}

\usepackage{graphicx,amsmath,tabularx}
\usepackage{color} 
\newcommand{\av}{_{\rm av}}
\newcommand{\smfrac}[2]{\mbox{\small $#1 \over #2$}}

\begin{document}

\title{Spin glasses in the non-extensive regime}
\author{Matthew Wittmann}
\author{A.~P.~Young}
\affiliation{Department of Physics, University of California, Santa Cruz, California 95064}

\date{\today}

\begin{abstract}
Spin systems with long-range interactions are ``non-extensive'' if the
strength of the interactions falls off sufficiently slowly with distance.  It
has been conjectured for ferromagnets, and more recently for
spin glasses, that, everywhere in the non-extensive regime, the
free energy is exactly equal to that for the infinite range model in which the
characteristic strength of the interaction is independent of distance.

In this paper we present the results of Monte Carlo simulations of the
one-dimensional long-range spin glasses in the
non-extensive regime. Using finite-size scaling, our results for the
transition temperatures are consistent with this prediction.  We also propose,
and provide numerical evidence for, an analogous result for \textit{diluted}
long-range spin glasses in which the coordination number is finite, namely
that the transition temperature throughout the non-extensive regime is equal
to that of the infinite-range model known as the Viana-Bray model.

\end{abstract}

\maketitle

\section{Introduction}
\label{sec:intro}

In the theory of phase transitions, it is often helpful to study models in a
range of dimensions from above the ``upper critical dimension'',
$d_u$, where mean-field critical behavior is valid, to below the ``lower
critical dimension'', $d_l$, where fluctuations destroy the transition.  For
Ising spin glasses $d_l \simeq 2.5$~\cite{boettcher:05} and $d_u = 6$.
However, it has been
difficult to cover this broad range
numerically for spin glasses, since $d_u$ is
quite large, and slow dynamics prevents equilibration at low temperatures
when the number of spins $N\ (= L^d)$ is greater than a few thousand.
It follows that at and above
$d_u$, one cannot study a sufficient range of linear sizes $L$
to perform the necessary
finite-size scaling (FSS) analysis.

As a result, there has been a lot of recent attention on long-range models in
one dimension, in which the interactions fall off with a power of the
distance. Such models have a long history going back to
Dyson~\cite{dyson:69,dyson:71}, who considered a ferromagnet with 
interactions $J_{ij}$
falling off like $1/r^\sigma$, and found a paramagnet-ferromagnet transition
for $\sigma \leq 2$. Kotliar et al.~\cite{kotliar:83} were the first to
study the spin
glass version of this model, which has received a lot of attention numerically
in the last few
years~\cite{katzgraber:05,katzgraber:09,leuzzi:08,moore:10,leuzzi:09}.

Varying the
power $\sigma$ in the
long-range spin glass model,
one has a range of behavior similar that obtained by varying
the dimension in short-range models, namely there is a ``lower critical
value'', $\sigma_l = 1$,~\cite{bray:86b}
\textit{above} which there is no transition at finite
temperature, and an ``upper critical value'', $\sigma_u =
2/3$~\cite{kotliar:83},
\textit{below} which
the transition has mean field critical exponents. Note that
\textit{increasing} $\sigma$ makes the interactions more short range, and so
corresponds to \textit{decreasing} $d$.

A precise connection between $d$ for short-range models and $\sigma$ for
long-range models can be made in the mean field region ($d >
6$ or $ 1/2 < \sigma < 2/3$), namely~\cite{larson:10}
\begin{equation}
d = {2 \over 2 \sigma - 1}, \qquad (d > 6,\ \  1/2 < \sigma < 2/3) \, .
\label{MF}
\end{equation}
This mapping 
shows that $d \to \infty$ for $\sigma \to 1/2$. Since the
transition temperature in mean field theory is given by
\begin{equation}
\left(T_c^{MF}\right)^2 = \sum_j \left[J_{ij}^2\right]\av, 
\label{TcMF}
\end{equation}
we see that for smaller values of $\sigma$,
i.e.\ $0 \le \sigma \le 1/2$, the strength of the interactions has to be
scaled with an inverse power of the system size to obtain a sensible
thermodynamic limit. We call this regime ``non-extensive''. The extreme
limit of this region, 
$\sigma = 0$, is the Sherrington Kirkpatrick (SK)
model~\cite{sherrington:75}, which is
``infinite-range''.  To complete the picture of the 1-d long-range spin glass
model, in this paper we study the non-extensive regime ($0 \le \sigma < 1/2$),
which has not been studied before, to our knowledge, apart from the SK model
($\sigma = 0$).

The non-extensive regime for ferromagnets has already been
investigated~\cite{cannas:00,campa:00}. This work shows that the behavior in
the whole non-extensive regime is the same, with a suitable rescaling of the
interactions, as that of the infinite-range ferromagnet in which every spin
interacts equally with every other spin, i.e.\ $\sigma = 0$. We give 
intuitive arguments for this in Appendix \ref{sec:spherical}.

It is interesting to ask if the same is true for spin glasses.
In a recent paper Mori~\cite{mori:11} has claimed that this is so, i.e.~for
all $0 \le \sigma < 1/2$ the behavior is the same as that of the SK model
($\sigma = 0$) provided the interactions are scaled with system size so that
$\sum_{j\ne i} [J_{ij}^2]\av $ is set to the same value for all $\sigma$.
However, this argument is just at the level of replicating the Hamiltonian so
it becomes translationally invariant, and then arguing that the earlier work
for ferromagnets can be taken over directly to prove the result.  While
plausible, this result is by no means rigorous and so we test it
here by Monte Carlo simulations.

One of the models we simulate here is the
usual one in which every spin interacts
with every other spin. However, it is also interesting to carry out the same
study for a diluted model~\cite{leuzzi:08} with a fixed average coordination
number $z$.  This model has received a lot of attention recently because the
computer time per sweep only varies as $N z$ 
(rather than $N^2$ for the undiluted model),
so it can be simulated much more efficiently than the
undiluted model for large $N$.  The diluted model with $\sigma = 0$ is called
the Viana-Bray~\cite{viana:85} model. It corresponds to a spin glass on a
random graph, the exact solution of which is expected to be the Bethe-Peierls
approximation.
By analogy with Mori's proposal, we
suggest here that the behavior of the
diluted spin glass model is identical to that of the Viana-Bray model $\sigma = 0$
everywhere in the non-extensive region ($0 \le \sigma < 1/2$).
We shall also provide
numerical evidence for this.

We should emphasize that \textit{universal} quantities, such as critical
exponents, are expected to be the same everywhere both in the mean field ($1/2 <
\sigma < 2/3$) and non-extensive ($0 \le \sigma < 1/2$) regimes. The claim
that we test is that \textit{all} the behavior of these models (not
just the critical behavior) is \textit{identical} for
all $\sigma$ in the non-extensive regime, at least in the thermodynamic limit.
We therefore need to look at \textit{non-universal}
quantities, and focus here on one particularly convenient quantity, the value
of the transition temperature $T_c$.

The plan of this paper is a follows: In Sec.~\ref{sec:models} we describe the
models used in the simulations and give their corresponding mean-field
transition temperatures. In Sec.~\ref{sec:method} we give the details of the
Monte Carlo simulations and FSS analysis. The results are given in
Sec.~\ref{sec:results} and our conclusions are summarized in
Sec.~\ref{sec:conclusions}. Appendix \ref{sec:spherical} provides an intuitive
explanation of why the behavior of the ferromagnet is independent of $\sigma$
in the non-extensive regime.

\section{Models}
\label{sec:models}

The Hamiltonian that we study is 
\begin{equation}
\mathcal{H} = - \sum_{\langle i, j\rangle} J_{ij} S_i S_j\, ,
\label{ham}
\end{equation}
where the $S_i,\ (i = 1, 2, \cdots, L)$
are Ising spins which take values $\pm 1$, and the $J_{ij}$
are statistically independent, quenched, random variables. The mean is taken
to be zero and the variance satisfies
\begin{equation}
\left[ J_{ij}^2 \right]\av \propto r_{ij}^{-2\sigma} \, ,
\end{equation}
where, for the distance $r_{ij}$ we put the sites on a ring and take the chord
distance between sites $i$ and $j$~\cite{katzgraber:03}, i.e.
\begin{equation}
r_{ij} = {L \over \pi} \sin\left({\pi |i-j| \over L}\right) \, .
\end{equation}

The form of the distribution of the $J_{ij}$ is different for the undiluted
and diluted models. For the undiluted case the distribution of the
$J_{ij}$ is Gaussian,
\begin{equation}
P\left(J_{ij}\right) = {1 \over \sqrt{2 \pi}\, \Delta J_{ij}} \, \exp\left(-J_{ij}^2 \over
2 \left(\Delta J_{ij}\right)^2 \right) \, , \ (\text{undiluted}),
\end{equation}
where the variance is given by
\begin{equation}
\left(\Delta J_{ij}\right)^2 
= { C^2 \over r_{ij}^{2\sigma} } \, ,
\label{C}
\end{equation}
in which $C$ is a constant to be determined below.

In order to compare models with different values of $\sigma$, for each
$\sigma$ and $L$, we scale the variance so that 
\begin{equation}
\sum_{j} \left[J_{ij}^2\right]\av = 1\, , \qquad \text{(undiluted)},
\label{constraint}
\end{equation}
where the sum is for fixed $i$ and we have $J_{ii} = 0$.
Equation~\eqref{constraint} determines the value of $C$ in Eq.~\eqref{C}.
Because we 
consider the non-extensive regime, $C$ must vanish for $L \to\infty$ like
$L^{\sigma -\smfrac{1}{2}}$. 

The expression for the mean-field transition temperature in Eq.~\eqref{TcMF}
is the exact result for the SK model, $\sigma = 0$. Hence, from
Eq.~\eqref{constraint}, we have
\begin{equation}
T_c(\sigma = 0) = 1\, \qquad \text{(undiluted)} \, .
\end{equation}

For the diluted model, rather than the \textit{strength} of the interaction
falling off like $1/r_{ij}^\sigma$, most bonds are absent and it is the
\textit{probability} of there being a non-zero bond which falls of with
distance (asymptotically 
like $1/r_{ij}^{2\sigma}$). If a bond is present it is chosen
from a Gaussian
distribution with mean zero and
variance unity (i.e.\ independent of $r_{ij}$). 
In other words
\begin{equation}
P\left(J_{ij}\right) = (1 - p_{ij}) \, \delta\left(J_{ij}\right) + p_{ij}\,
{1 \over \sqrt{2 \pi}} \, e^{-J_{ij}^2/2} 
\, , \ \ (\text{diluted}), 
\label{P_dil}
\end{equation}
where $p_{ij} \propto 1/r_{ij}^{2\sigma}$ at large distance. 

It is convenient to
fix the mean number of neighbors $z$. The pairs of sites with non-zero bonds
are then generated as follows. Pick a site $i$ at random. Then pick a site $j$
with probability $\widetilde{p}_{ij} = A/ r_{ij}^{2\sigma}$,
where $A$ is 
determined by normalization.
If there is already a bond between $i$ and $j$
repeat until
a pair $i, j$ is selected which does not already have a
bond\footnote{Note that if 
$z \widetilde{p}_{ij} \ll 1$ then $p_{ij}$ in Eq.~\eqref{P_dil} is
given by $p_{ij} = z \widetilde{p}_{ij}$, but
otherwise there are corrections due to rejection of pairs $i, j$ when there is
already a bond between them.}.
At that point set $J_{ij}$ equal
to a Gaussian random variable with zero mean and variance unity. This process
is repeated $N z/2$ times so the number of sites connected to a given site has
a Poisson distribution with mean $z$. Because each site has, on average, $z$
neighbors, and the variance of each interaction is unity, we have
\begin{equation}
\sum_{j} \left[J_{ij}^2\right]\av = z\, , \qquad \text{(diluted)}\, .
\label{constraint_dil}
\end{equation}

The transition temperature for the diluted model with $\sigma = 0$ was shown
by Viana and Bray~\cite{viana:85} to be given by the solution of
\begin{equation}
{1 \over \sqrt{2 \pi}}\int_{-\infty}^\infty d x\, e^{-x^2/2}\,
\tanh^2\left({x \over T_c}\right) = {1 \over z}.
\label{TcVB}
\end{equation}
We choose $z = 6$ for which we find
\begin{equation}
T_c(\sigma = 0) = 2.0564\, ,\quad \text{(diluted)} \, .
\label{TcVBz6}
\end{equation}

\section{\label{sec:method}Method}

\begin{table}[!tb]
\begin{center}
\begin{tabular*}{\columnwidth}{@{\extracolsep{\fill}} ||r|r|r|r|r|r|r|r||}
\hline\hline
$\sigma$ & $L$  &$N_\text{samp}$& $N_\text{equil}$ & $N_\text{meas}$ & $T_\text{min}$ & $T_\text{max}$ & $N_T$ \\
\hline
     0 &  64 & 16000   &    1000   &   10000 &   0.5 &  1.65 &  47 \\
     0 & 128 & 16000   &    1000   &   10000 &   0.5 &   1.6 &  45 \\
     0 & 256 & 16000   &    1000   &   10000 &   0.5 &   1.6 &  45 \\
     0 & 512 &  8000   &    1000   &   10000 &  0.75 &  1.55 &  33 \\
     0 &1024 &  8000   &    1000   &   10000 &  0.75 &   1.5 &  31 \\
     0 &2048 &  4000   &    1000   &   10000 &  0.75 &   1.5 &  31 \\
     0 &4096 &  4000   &    2000   &   10000 &  0.85 & 1.525 &  28 \\
  0.25 &  64 & 16000   &    1000   &   10000 &   0.5 &  1.65 &  47 \\
  0.25 & 128 & 16000   &    1000   &   10000 &   0.5 &   1.6 &  45 \\
  0.25 & 256 & 16000   &    1000   &   10000 &   0.5 &   1.6 &  45 \\
  0.25 & 512 &  8000   &    1000   &   10000 &   0.5 & 1.525 &  42 \\
  0.25 &1024 &  8000   &    1000   &   10000 &  0.75 &   1.5 &  31 \\
  0.25 &2048 &  4000   &    1000   &   10000 &  0.75 &   1.5 &  31 \\
  0.25 &4096 &  4000   &    2000   &   10000 &  0.85 & 1.525 &  28 \\
\hline\hline
\end{tabular*}
\end{center}
\caption{
Simulation parameters for the undiluted models. $N_\text{samp}$ is the number
of samples, $N_\text{equil}$ and $N_\text{meas}$ are the number of sweeps for
equilibration and for the measurement phase, respectively. We simulate $N_T$
temperatures between $T_\text{min}$ and $T_\text{max}$.}
\label{tab:params_undil}
\end{table}
\begin{table}[!tb]
\begin{center}
\begin{tabular*}{\columnwidth}{@{\extracolsep{\fill}} ||r|r|r|r|r|r|r|r||}
\hline\hline
$\sigma$ & $L$  &$N_\text{samp}$& $N_\text{equil}$ & $N_\text{meas}$ & $T_\text{min}$ & $T_\text{max}$ & $N_T$ \\
\hline
     0 &   256  & 8000   &     400   &    8000 &  1.85  &  2.5 &  27 \\
     0 &   512  & 8000   &     800   &   16000 &  1.85  &  2.5 &  27 \\
     0 &  1024  & 8000   &    2000   &   40000 &  1.85  &  2.5 &  27 \\
     0 &  2048  & 4000   &    2000   &   40000 &  1.85  &  2.5 &  27 \\
     0 &  4096  & 4000   &    2000   &   40000 &   1.9  &  2.5 &  25 \\
     0 &  8192  & 2000   &    4000   &   80000 &   1.9  &  2.5 &  25 \\
     0 & 16384  & 2000   &    4000   &   80000 &   2.0  &  2.5 &  14 \\
  0.25 &   256  & 8000   &     800   &   16000 &  1.85  &  2.5 &  27 \\
  0.25 &   512  & 8000   &     800   &   16000 &  1.85  &  2.5 &  27 \\
  0.25 &  1024  & 8000   &    1200   &   24000 &  1.85  &  2.5 &  27 \\
  0.25 &  2048  & 4000   &    2000   &   40000 &  1.85  &  2.5 &  27 \\
  0.25 &  4096  & 4000   &    2000   &   40000 &   1.9  &  2.5 &  25 \\
  0.25 &  8192  & 2000   &    4000   &   80000 &   1.9  &  2.5 &  25 \\
  0.25 & 16384  & 2000   &    4000   &   80000 &   2.0  &  2.5 &  14 \\
 0.375 &   256  &32000   &    1200   &   24000 & 1.863  &  4.0 &  24 \\
 0.375 &   512  &26327   &    1200   &   24000 & 1.863  &  4.0 &  26 \\
 0.375 &  1024  &16000   &    1200   &   24000 & 1.913  &  4.0 &  24 \\
 0.375 &  2048  &15998   &    2000   &   40000 &  1.95  &  4.0 &  24 \\
 0.375 &  4096  & 8000   &    4000   &   80000 & 1.962  &  4.0 &  28 \\
 0.375 &  8192  & 7999   &    4000   &   80000 & 1.975  &  4.0 &  34 \\
 0.375 & 16384  & 4000   &    4000   &   80000 &   2.0  & 2.51 &  18 \\
\hline\hline
\end{tabular*}
\end{center}
\caption{
Simulation parameters for the diluted models. The parameters are the same as
in Table \ref{tab:params_undil}}.
\label{tab:params_dil}
\end{table}

We perform Monte Carlo simulations on the models described in
Sec.~\ref{sec:models}.
To speed up equilibration we use the parallel tempering (exchange)
Monte Carlo method \cite{hukushima:96}. In this approach one simulates
$N_T$ copies of the spins with the same interactions, each at a
different temperature between a minimum value $T_\text{min}$ and a
maximum value $T_\text{max}$. In addition to the usual single spin-flip
moves for each copy, we perform global moves in which we interchange
the temperatures of two copies at neighboring temperatures with a
probability which satisfies the detailed balance condition.  In this
way, the temperature of a particular copy performs a random walk
between $T_\text{min}$ and $T_\text{max}$, thus helping to overcome
the free energy barriers found in the simulation of glassy systems.

For the simulations of the undiluted model to be in equilibrium 
the following equality must be satisfied~\cite{katzgraber:03},
\begin{equation}
U = - \frac{\left(T_c^{MF}\right)^2}{2 T}\, (1 - q_l)\, ,
\qquad \text{(undiluted)},
\label{equil}
\end{equation}
where
\begin{equation}
U = -\sum_{\langle i, j \rangle} \left[
J_{ij} \langle S_i S_j \rangle \right]\av \, ,
\end{equation}
is the average energy, and
$q_l$ is the ``link overlap'' defined by
\begin{equation}
q_l = {2 \over N}\sum_{\langle i,j\rangle } {[J_{ij}^2]_{\rm av} \over
(T_c^{MF})^2} 
[ \langle S_i S_j \rangle^2 ]_{\rm av} , \qquad\text{(undiluted)}.
\label{ql}
\end{equation}
in which $T_c^{MF}$ is given by Eq.~\eqref{TcMF} (and here
set equal to unity by
the scaling of the interactions, see Eq.~\eqref{constraint}).
Equation \eqref{equil} is obtained by integrating by parts with respect to the
$J_{ij}$ the expression for the average energy, and noting that the
distribution is Gaussian. This equation is useful because, very plausibly, the two sides
approach their common value from \textit{opposite}
directions~\cite{katzgraber:03},
so, if the two sides agree,
the system has reached equilibrium (at least for the energy
and link overlap).

For the diluted model, the equilibration test takes the 
form,~\cite{katzgraber:05}
\begin{equation}
U = - \frac{z}{2 T}\,(1 - q_l) \, , \qquad \text{(diluted)},
\label{equil_dil}
\end{equation}
where the link overlap
is now defined by 
\begin{equation}
q_l = {2\over N z} \,\sum_{\langle i, j \rangle}
\left[\epsilon_{ij} \langle S_i S_j \rangle^2\rangle\right]\av, \qquad
\text{(diluted)},
\end{equation}
in which $\epsilon_{ij} = 1$ if there is a bond between $i$ and $j$,
and zero otherwise. As with Eq.~\eqref{equil}, we expect that the two sides of
Eq.~\eqref{equil_dil} approach each other from opposite directions as
equilibrium is approached.

We consider results obtained by
successively doubling the number of sweeps, in each case averaging over
the last half of the sweeps, and we accept the data as being in equilibrium if the
last three data points agree with each other within the
error bars. The total number of sweeps used in this check is shown as
$N_\text{equil}$ in Tables
\ref{tab:params_undil} and \ref{tab:params_dil}. We then do ``production" runs
where, in addition to $N_\text{equil}$ sweeps for equilibration, we do
10 to 20 times as many sweeps,
$N_\text{meas}$, during which measurements are performed.
All the parameters used in the simulations are given in Tables
\ref{tab:params_undil} and \ref{tab:params_dil}.
To avoid bias,
each distinct thermal average, for example in Eq.~\eqref{ql},
is evaluated in a separate copy (replica)
of the system with the same interactions.

\begin{figure*}[!tb]
\begin{center}
\includegraphics[width=\figurewidth]{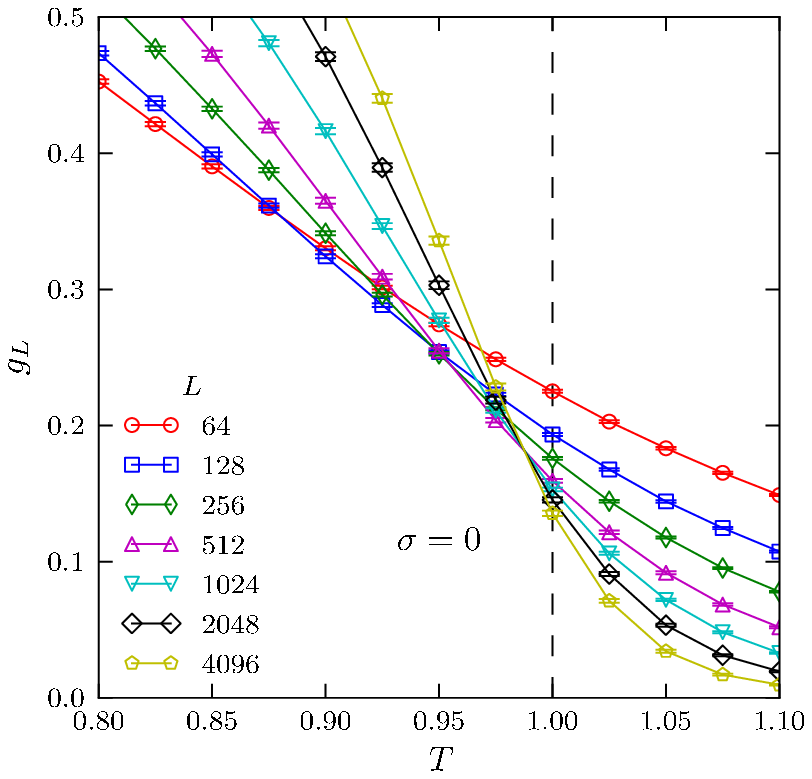}
\includegraphics[width=\figurewidth]{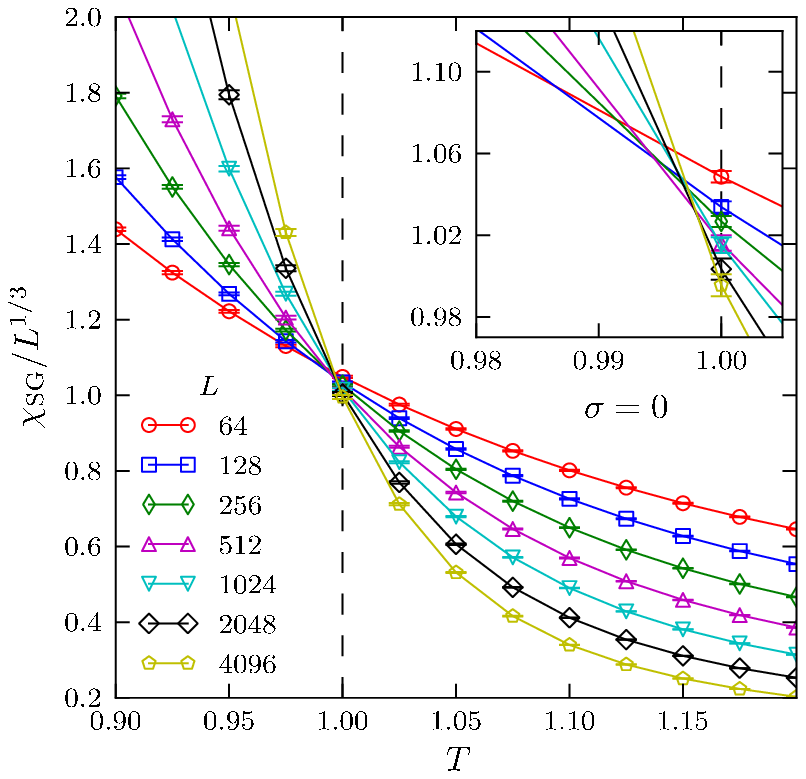}
\includegraphics[width=\figurewidth]{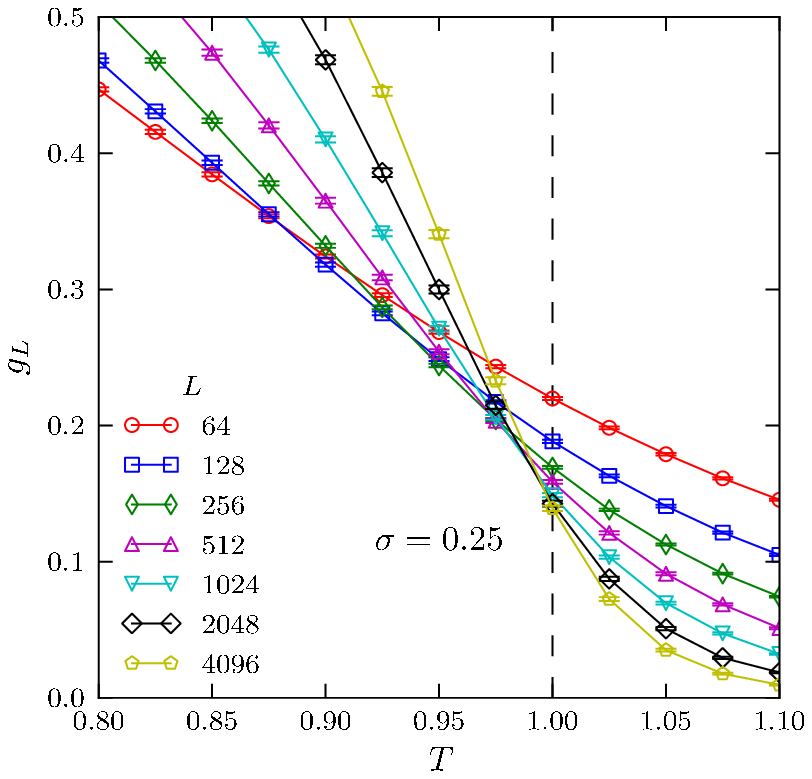}
\includegraphics[width=\figurewidth]{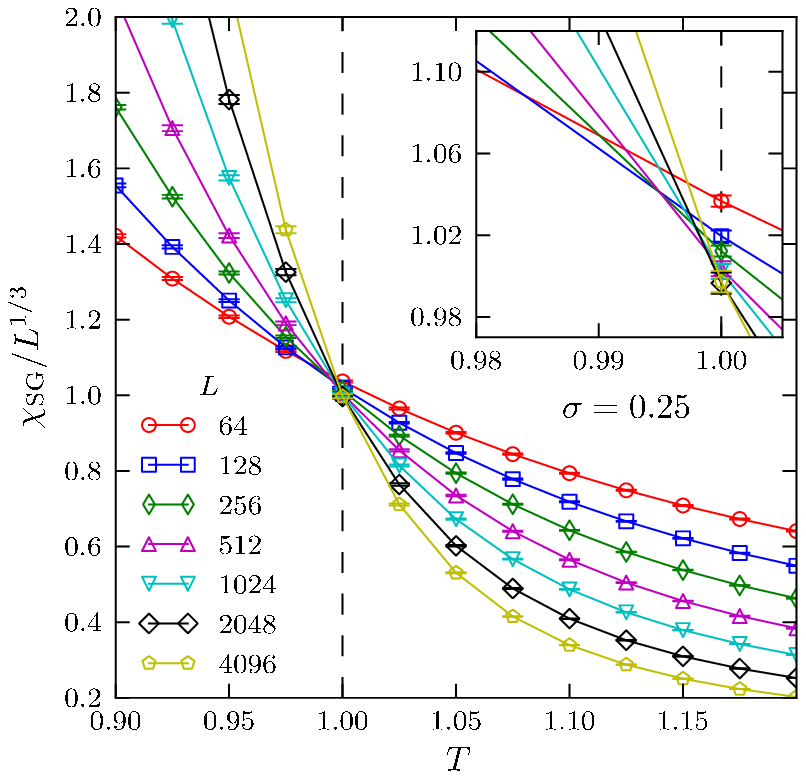}
\caption{(Color online) 
Data for the the scaled spin glass susceptibility and the Binder ratio
for the undiluted model with $\sigma = 0$, the SK model,
(top), and the undiluted model with $\sigma =
0.25$ (bottom). The dashed vertical line shows the exact value of the
transition temperature ($T_c = 1$) for the SK model.
\label{fig:undil_data}}
\end{center}
\end{figure*}
\begin{figure*}[!tb]
\begin{center}
\includegraphics[width=\figurewidth]{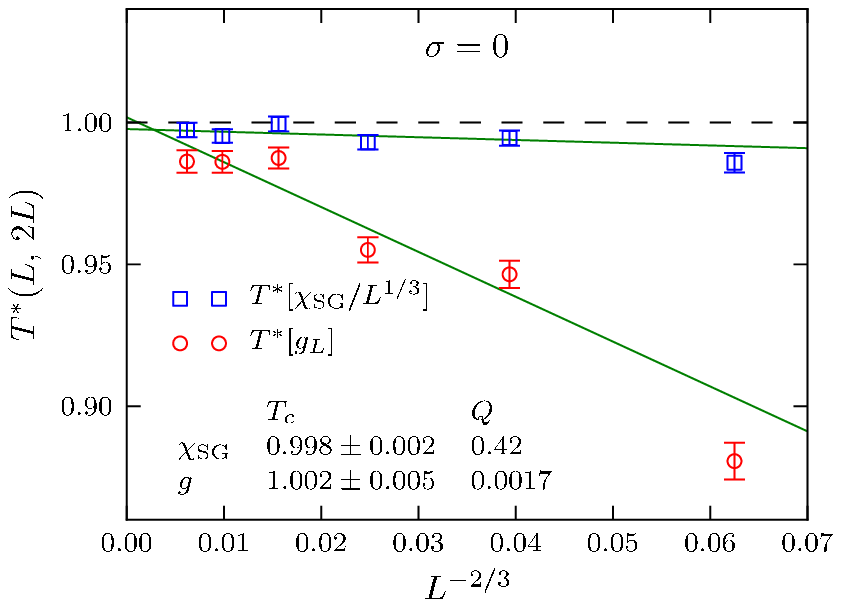}
\includegraphics[width=\figurewidth]{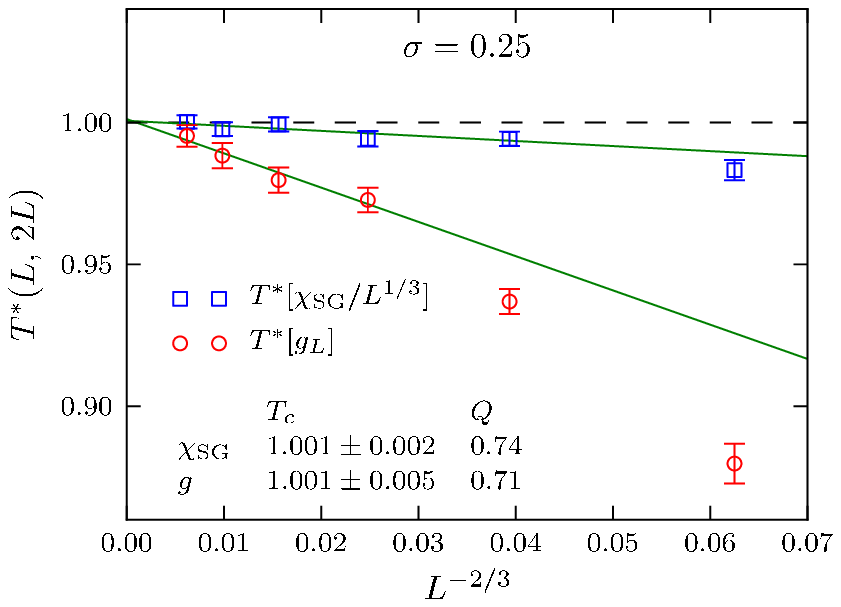}
\caption{(Color online) 
Results for the intersection temperatures for the SK model (left) and the
undiluted $\sigma = 0.25$ model (right).
\label{fig:undil_Tstar_0_0.25}}
\end{center}
\end{figure*}

We focus on moments of the spin glass order parameter $q$ where
\begin{equation}
q = {1 \over L} \,\sum_{i} S_i^{(1)} S_i^{(2)} \, ,
\end{equation}
in which ``$(1)$'' and ``$(2)$''
refer to two independent copies of the system with the
same interactions. Of particular interest are the spin glass susceptibility
\begin{equation}
\chi_{SG} = L \langle q^2 \rangle\, ,
\end{equation}
and the Binder ratio
\begin{equation}
g = {1 \over 2}\left(3 - {[\langle q^4 \rangle]\av \over 
[\langle q^2 \rangle]\av^2} \right) \, .
\end{equation}

\begin{figure*}[!tb]
\begin{center}
\includegraphics[width=\figurewidth]{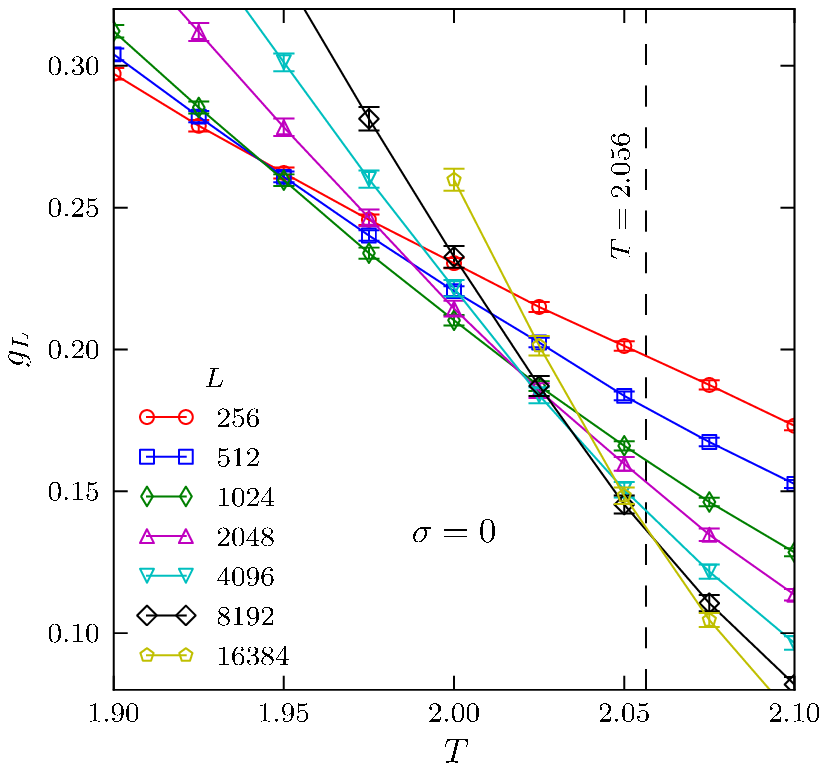}
\includegraphics[width=\figurewidth]{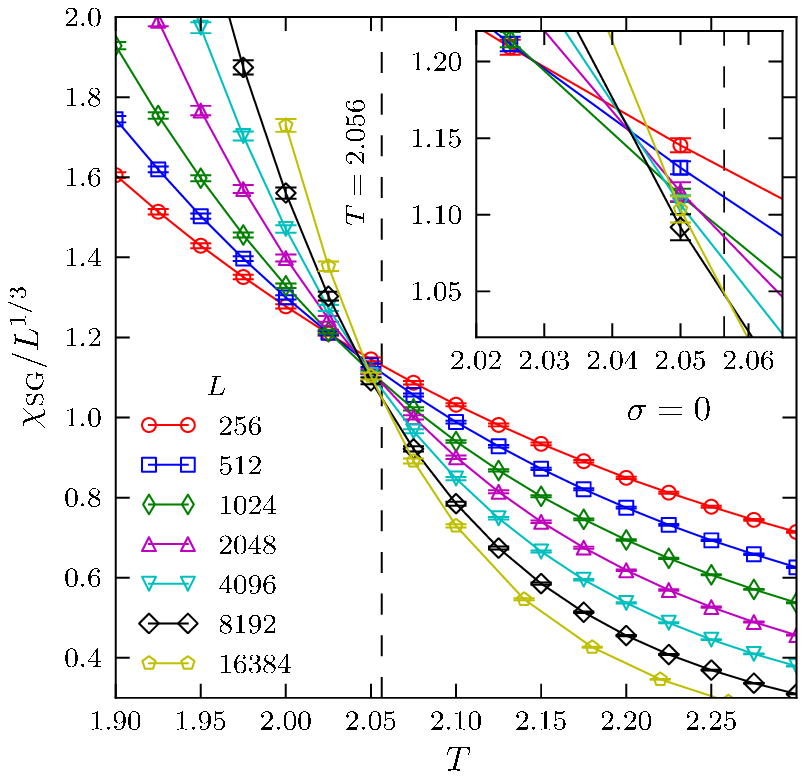}
\includegraphics[width=\figurewidth]{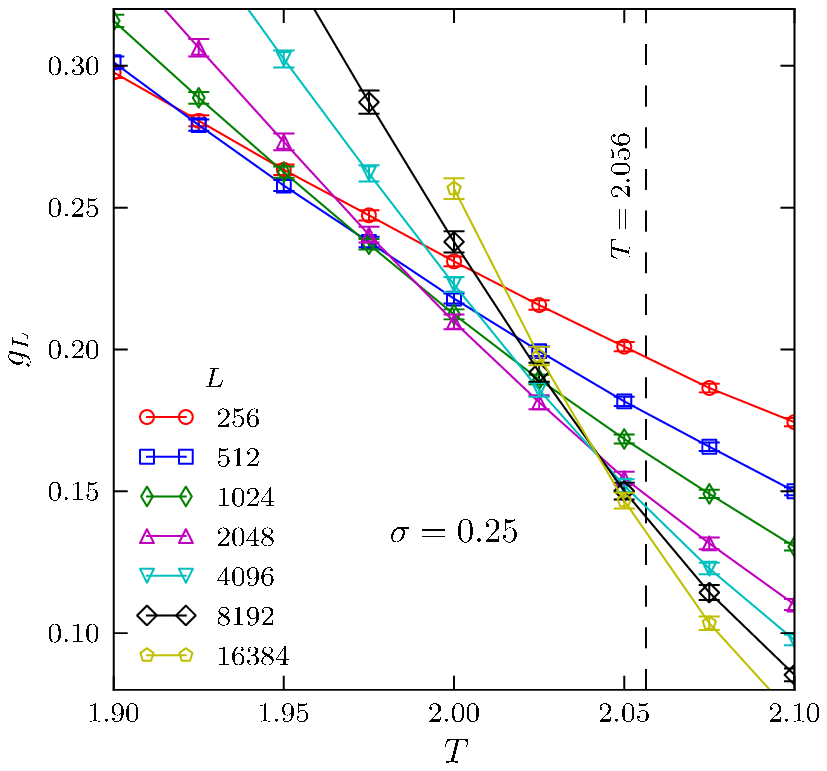}
\includegraphics[width=\figurewidth]{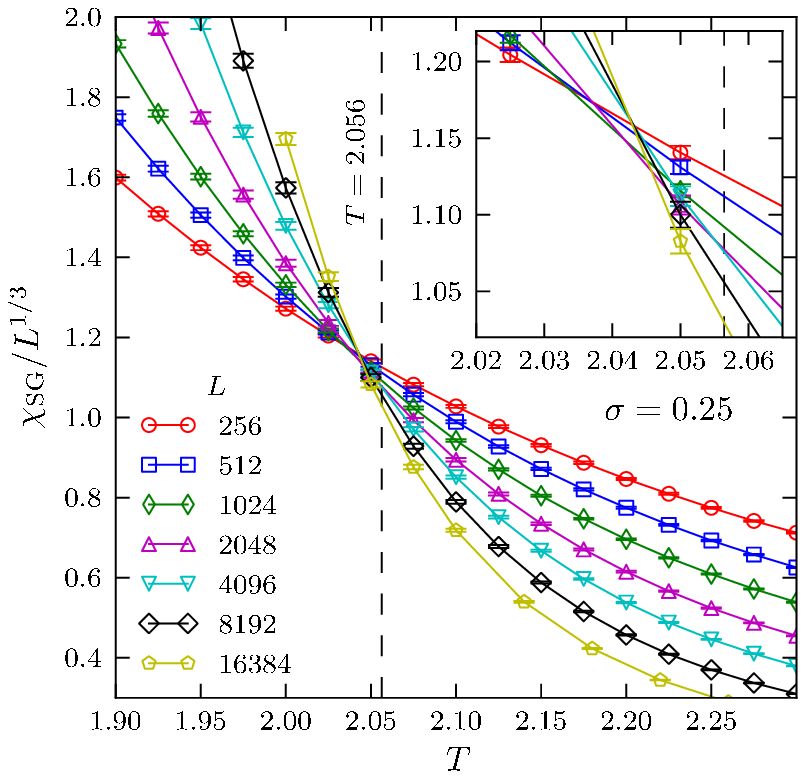}
\caption{(Color online) 
Data for the the scaled spin glass susceptibility and the Binder ratio
for the Viana-Bray model, i.e.\ the diluted model with
$\sigma = 0$ (top), and the diluted model with $\sigma = 0.25$ (bottom).
The dashed vertical line shows the transition temperature for the Viana-Bray
model obtained from Eq.~\eqref{TcVB} of the text.
\label{fig:dil_data}}
\end{center}
\end{figure*}
\begin{figure*}[!tb]
\begin{center}
\includegraphics[width=\figurewidth]{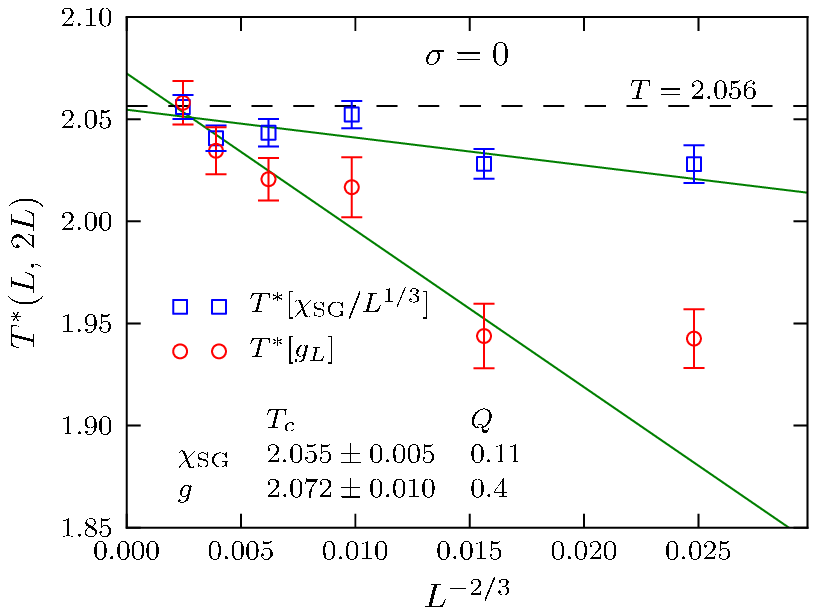}
\includegraphics[width=\figurewidth]{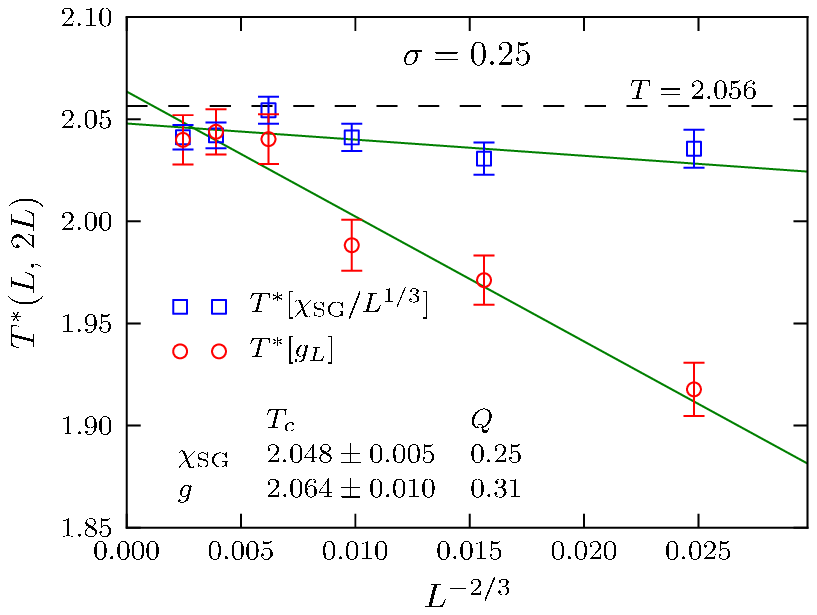}
\caption{(Color online) 
Results for the intersection temperatures for the Viana-Bray model (left) and 
the diluted model with $\sigma = 0.25$ (right).
\label{fig:dil_Tstar_0_0.25}}
\end{center}
\end{figure*}

Since the Binder ratio is dimensionless, its finite-size scaling (FSS)
behavior is
simple. We are always in the regime of mean field critical exponents
($0 \le \sigma < 2/3$), so it has the form~\cite{fss:gtlcd}
\begin{equation}
g = \widetilde{g} \left(\, (T - T_c)\, L^{1/3} \right) \, .
\label{g_FSS}
\end{equation}
The spin glass susceptibility is not dimensionless but, since we are in the
mean field regime, its FSS form is also known exactly. It has
the form~\cite{fss:gtlcd}
\begin{equation}
\chi_{SG} = L^{1/3}\, \widetilde{\chi} \left(\, (T - T_c)\, L^{1/3} \right) \, .
\label{chisg_FSS}
\end{equation}

One can therefore determine the transition temperature from where
the data for $g$
or $\chi_{SG}/L^{1/3}$ for different sizes intersects. However, we shall see
that the data does not all intersect at a single temperature, showing 
that there are
corrections to the FSS form in Eqs.~\eqref{g_FSS} and \eqref{chisg_FSS}.
Consider Eq.~\eqref{chisg_FSS}. According to standard finite-size scaling, the
spin-glass
susceptibility normally varies near the critical point according
to~\cite{larson:10}
\begin{equation}
\chi_{SG}(t, L) = L^a\left [ f(L^b t) + L^{-\omega} g(L^y t) + 
\cdots \right] + c_0 + c_1 t + \cdots ,
\label{fss}
\end{equation}
where $t = T - T_c$, $a = 2 - \eta \ (= 2\sigma -1\ \text{here})$, and $b = 1 /\nu$.
The $L^{-\omega}$ term is the leading
\textit{singular} correction to scaling and $c_0$ is the leading
\textit{analytic} correction to scaling. However, in the mean field limit,
$\sigma < 2/3$, the exponents $a$ and $b$ are
independent of $\sigma$\cite{binder:85,luijten:99,jones:05}
and take the value at $\sigma_c$ for all $1/2 < \sigma <
\sigma_c$, i.e. $a=b= 1/3$.  Furthermore, although the $L^{2\sigma
-1}$ term is replaced as the \textit{largest} term by an
$L^{1/3}$ term (due to the presence of a ``dangerous irrelevant
variable,''cf.~Refs.~[\onlinecite{binder:85,luijten:99,jones:05}])
we expect~\cite{larson:10} this term to not disappear but rather become a
\textit{correction to scaling}. Hence, we replace Eq.~\eqref{fss} by
\begin{multline}
\chi_{SG}(t, L) = L^{1/3}\left [ f(L^{1/3} t) + L^{-\omega} g(L^{1/3} t) +
\cdots \right] \\
+ d_0 L^{2\sigma-1} hg(L^{1/3} t) + c_0 + c_1 t \cdots .
\label{fss2}
\end{multline}
The correction exponent $\omega$ can be obtained in the mean-field
regime from the work of Kotliar {\em et al}.~\cite{kotliar:83} and
is given by $\omega = 2 - 3 \sigma$. Hence, in the non-extensive regime,
$\sigma < 1/2$ the dominant correction to scaling is the constant $c_0$.


Adding a constant to the RHS of Eq.~\eqref{chisg_FSS} 
it is straightforward to show
that the intersection temperature of the data for $\chi_{SG}/L^{1/3}$ for
sizes $L$ and $2L$ is given by
\begin{equation}
T^\star(L, 2L) = T_c + {A \over L^{2/3}}+ \cdots \, ,
\label{Tstar}
\end{equation}
where $A$ is a constant and the omitted terms are higher order in $1/L$.
We expect that the intersection temperatures for the
data for $g$ have the same form. We shall use Eq.~\eqref{Tstar} to determine
$T_c$ for the models studied.

\section{Results}
\label{sec:results}
We first present our results for the undiluted model. Data for the the scaled
spin glass susceptibility and the Binder ratio are shown in
Fig.~\ref{fig:undil_data}. The top part is 
for the SK model, $\sigma = 0$, and the bottom part is for the 
undiluted model with $\sigma =
0.25$. One sees large corrections to scaling for the Binder ratio (the
left-hand figures) but much smaller corrections for the scaled spin glass
susceptibility (the right-hand figures).
The inset enlarges the region of the intersections for the
latter data.

Figure~\ref{fig:undil_Tstar_0_0.25}
shows values for the intersection temperature. These were
determined by 
interpolation using cubic splines,
and error bars computed by a jackknife analysis.
For \textit{both}
values of $\sigma$ the data extrapolates to a value of 1,
the exact value for the SK model,
(with very small errors). The quality of the fit, as represented by the
goodness of fit parameter $Q$~\cite{press:07},
is satisfactory except for the Binder ratio data for
the SK model. We don't have a good explanation for this, except perhaps that 
multiple corrections to scaling are significant for the range of sizes
studied. In any case we note that the result $T_c = 1$ for the SK model is
rigorously correct.
The result that $T_c = 1$ also for $\sigma = 0.25$, at the midpoint of the
non-extensive region, provides strong evidence for the claim of
Mori~\cite{mori:11} that all models in the non-extensive region are identical
to the SK model. While it would be useful to check this also in the space
glass phase below $T_c$, such simulations would be difficult because relaxation
times increase dramatically at low $T$ and so the range of sizes that could be
studied would be much more limited than in the data presented here.

\begin{figure}[!tb]
\begin{center}
\includegraphics[width=\figurewidth]{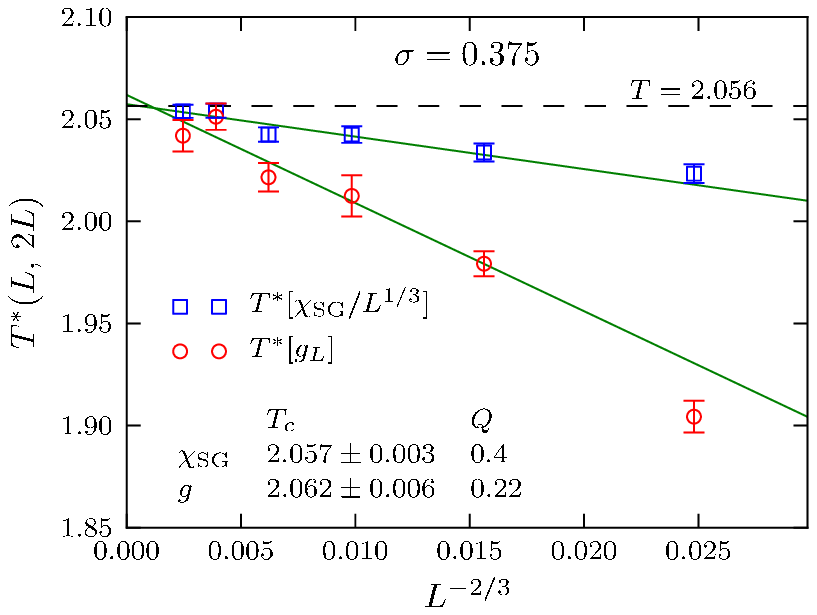}
\caption{(Color online) 
Results for the intersection temperatures for the 
diluted model with $\sigma = 0.375$.
\label{fig:dil_Tstar_0.375}}
\end{center}
\end{figure}

The corresponding results for the diluted model for $\sigma = 0$ and $0.25$
are shown in
Figs.~\ref{fig:dil_data} and \ref{fig:dil_Tstar_0_0.25}. We also performed
simulations for $\sigma = 0.375$ and show the resulting intersection
temperatures in Fig.~\ref{fig:dil_Tstar_0.375}. For $\sigma = 0$, the
Viana-Bray~\cite{viana:85} model, the transition temperature is given by
Eq.~\eqref{TcVB} which, for $z=6$ taken here, gives the result in
Eq.~\eqref{TcVBz6}. In Fig.~\ref{fig:dil_data} we again see that corrections
to scaling are larger for the Binder ratio than for the scaled spin glass
susceptibility.  The intersection temperatures \textit{all} extrapolate to the
exact value for $\sigma = 0$ within statistical uncertainty\footnote{All the
results within one standard deviation except for the data for $g$ for $\sigma
= 0$ and $\chi_{SG}/L^{1/3}$ for $\sigma = 0.25$ but even these are within
$\sim 1.5$ standard deviations, which we also consider acceptable.}.

\section{Summary and Conclusions}
\label{sec:conclusions}
We have performed Monte Carlo simulations to investigate the transition
temperature
of one-dimensional Ising spin glasses, both undiluted and
diluted, for several values of $\sigma$ in the non-extensive regime
$0 \le \sigma < 1/2$.
For the undiluted model we studied two values of $\sigma$, $\sigma=0$ and
$\sigma=0.25$. For $\sigma=0.25$, which lies in the middle of the non-extensive
regime, we find that the transition temperature agrees to high precision with
the exact solution of the SK model. As a check, we also simulated the
$\sigma=0$ case, obtaining results consistent with the exact SK model result, 
though there seem to be multiple corrections to FSS for some of the data.

For the diluted model we studied three values of $\sigma$: $\sigma=0$, which
corresponds to the Viana-Bray model, $\sigma=0.25$, which lies in the middle of
the non-extensive regime, and $\sigma=0.375$. In all cases we found the
transition temperature to be consistent with the exact solution of the
Viana-Bray model; all results are within $\sim1.5$ standard deviations.

To conclude,
our results provide confirmation of the proposal~\cite{mori:11} that 
the behavior of (undiluted) spin glasses everywhere in the non-extensive regime
is identical to that of the SK model. We have also proposed that an
analogous result applies to \textit{diluted} spin glass models, and provided
numerical evidence for this too.

\begin{acknowledgments}
{
This work is supported in part by the National
Science Foundation under Grant
No.~DMR-0906366.
We would also like to thank the Hierarchical Systems Research
Foundation for generous provision of computer support.
}

\end{acknowledgments}

\appendix
\section{Spherical Approximation for the ferromagnet}
\label{sec:spherical}
For the infinite-range \textit{ferromagnet}, the interaction $J_{ij}$ is equal
to $1/(N-1)$ for $i \ne j$ and 0 for $i = j$. This fixes $T_c = 1$.
The Fourier transform of this interaction is given by
\begin{equation}
J(k) = \delta_{k, 0} - {1 \over N} \, ,
\label{Jk_ir}
\end{equation}
so only the $k=0$ mode contributes to the transition.

For a power-law decay of the interactions
in the non-extensive regime ($0 < \sigma < 1$),
on dimensional grounds
there is a singular piece
which diverges like $k^{\sigma - 1}$ for $k \to 0$. Furthermore the
interactions have to be multiplied by a number of order $N^{\sigma - 1}$ in
order to satisfy the condition $T_c^{MF} = \sum_j J_{ij} = 1$. Hence, roughly
speaking, we have
\begin{equation}
J(k) \propto \left(k\, N\right)^{\sigma - 1} \, ,
\label{Jk_lr}
\end{equation}
where we note that $k \equiv k_n =
2 \pi n / L, \ (n = 0, 1, 2, \cdots)$. (For $n=0, J(0)$
does not actually diverge but will be comparable to $J(k_1)$). Hence other
long wavelength modes, in addition to $k = 0$, are now significant. However,
we shall now see
that there are not enough of them to change the value of $T_c$ from that of
$T_c^{MF} \ (= 1)$.

We will do this by considering the ``spherical
approximation''~\cite{berlin:52}.
in which we reexpress the problem as a Gaussian one, with ``soft'' spins
$\phi_i$ which take values from $-\infty$ to $\infty$, and a Hamiltonian
\begin{equation}
\mathcal{H}_\text{Gauss} = {1 \over 2} \sum_{i,j}\, 
\left(\mu \delta_{ij} - J_{ij}
\right) \, \phi_i \phi_j \, ,
\end{equation}
where $\mu$ is a Lagrange multiplier whose value is chosen to enforce the
length constraint
\begin{equation}
\langle \phi_i^2 \rangle = 1 \, .
\label{length_constraint}
\end{equation}
It turns out the the spherical approximation is exact for an $m$-component
model in the limit $m \to \infty$~\cite{stanley:68}.
Fourier transforming Eq.~\eqref{length_constraint} and doing the Gaussian integrals gives
\begin{equation}
{1 \over T} = {1 \over L} \sum_k {1 \over \mu - J(k) } \, .
\end{equation}
The transition occurs when the denominator vanishes at $k= 0$, i.e.~when
$\mu = J(0)$ and so
\begin{equation}
{1 \over T_c^\text{spher}} = {1 \over L} \sum_k {1 \over J(0) - J(k) }
\label{Tc_spher}
\, .
\end{equation}
It is interesting to compare this with the mean field result, $T_c = \sum_j
J_{ij} = J(0)$. Since $J_{ii} = (1/L) \sum_k J(k) = 0$ we can rewrite the mean
field transition temperature as
\begin{equation}
T_c^{MF} = {1 \over L} \sum_k \left[\,J(0) - J(k)\,\right] \, .
\label{Tc_MF}
\end{equation}
Thus, whereas in mean field theory, $T_c$ is equal to the average of $J(0) -
J(k)$, in the spherical approximation $1/T_c$ is equal to the average of the
\textit{inverse} of this.

For the infinite range model, where $J(k)$ is given by Eq.~\eqref{Jk_ir} and
only the $k=0$ mode contributes, the spherical result agrees with the mean
field result (consistent with the MF result being exact for this model).

We now estimate $T_c$ from the spherical approximation, Eq.~\eqref{Tc_spher},
for
the power-law model, where $J(k)$ varies like Eq.~\eqref{Jk_lr}.
Because we normalize the interactions to $J(0)= 1$, we can include an extra factor
of $J(0)$ and expand in powers of $J(k) / J(0)$, i.e.
\begin{subequations}
\begin{align}
{1 \over T_c^\text{spher}} &= {1 \over L} \sum_k{ 1 \over 1 - J(k)/J(0)} \, \\
&= {1 \over L} \sum_k \left[1 + {J(k) \over J(0)} + \left(
{J(k) \over J(0)}\right)^2 + \cdots \right] \\
&= 1 + {1 \over L} \sum_k \left[ \left( {J(k) \over J(0)}\right)^2 + \cdots
\right] \label{ntolast}\\
&= 1 + {\sum_j J_{ij}^2 \over \left(\sum_j J_{ij} \right)^2} +
\cdots
\, ,
\label{last}
\end{align}
\end{subequations}
where in Eq.~\eqref{ntolast} we used that $\sum_k J(k) = 0$. In
Eq.~\eqref{last} we have $\sum_j J_{ij} \propto L^{1-\sigma}$ while
$\sum_j J_{ij}^2 = L^{1-2\sigma}\, (0 \le \sigma < 1/2), \sum_j J_{ij}^2 =
\text{const.}\, (1/2 < \sigma < 1)$. Hence the leading correction term in
Eq.~\eqref{last} vanishes everywhere in the non-extensive regime.

To conclude, in this appendix we have given a suggestive argument as to why $T_c$
for the ferromagnet is given
exactly by the mean field value everywhere in the non-extensive regime. It is
therefore also plausible that other properties are also identical to those of
mean field theory (i.e.~the infinite-range model.) 

\bibliography{refs,comments}

\begin{thebibliography}{27}
\bibitem[{fss()}]{fss:gtlcd}
\bibinfo{note}{For a discussion of how finite-size scaling is modified in the
  region of mean-field exponents, see for example
  Refs.~\cite{brezin:82,brezin:85,binder:85,luijten:99,jones:05}.}
\end{thebibliography}

\end{document}